# Research on PM emission and energy consumption of mixed traffic flow based on cellular automata FI model

Dong Chen[a*,] Yanping Huang[a], Xue Wang[b], Mi Tan[c], Bingling Cen[b]

[a] School of Management Science and Engineering, Guangxi University of Finance and Economics, Nanning, Guangxi 530007, China
[b] School of Physical Science and Technology, Guangxi University, Nanning 53004, China
[c] School of Business, Macau University of Science and Technology, Macau 999078, China

**Abstract**

Traffic congestion leads to a sharp increase in fuel consumption and exhaust emissions, which easily leads to air pollution and fuel waste. Based on Fukui-Ishibashi (FI) model, this paper studies particulate matter(PM) emission and energy dissipation of traffic flow by considering the maximum speed, mixing ratio and random slowing probability. Simulation results show that the energy dissipation of the traffic flow consisting of the same length and different maximum velocity is associated with the maximum velocity. The greater the maximum velocity, the more the energy consumes. When the maximum velocity is different, the PM emission is similar to the change of energy consumption. Moreover, for the same maximum velocity and different mixing ratio, the energy dissipation and the PM emission related to the mixing ratio. There are the maximum fuel consumption and emission value in the maximum flow. For the same maximum velocity and length of vehicles, their energy consumption and emission are determined by random slowing probability.

## 1. Introduction

The growth of population within major cities has resulted in an unprecedented increase in activities and higher demands for energy and transportation. These factors contribute significantly to urban air pollution emanating from often congested major road networks.[1] The emission of particulate matter (PM), NOx, $CO_2$, $SO_2$ and other harmful substances of vehicles even exceeded that of industrial pollutants, seriously endangering human health.[2] They were well known to be associated with classical, widespread occupational lung disease such as the pneumoconiosis, as well as with systemic intoxications such as lead poisoning, especially at higher levels of exposure.[3] Transportation energy is a significant portion of the national economy.[4] Pollutant emission has become a vexed question and one of the most challenging global problem for air quality mitigation and for climate change policies.[5] Vehicles emissions had unavoidable responsibility that cause hazy weather, respiratory diseases and traffic congestion.[6] Therefore it was the most pressing problem to improve city atmosphere quality by dealing with emissions form transportation.[7] Authorities have tried to improve air quality in metropolis by introducing strict emission standards, using cleaner energy and retrofitting catalytic converters and particle traps.[8] In addition, the study of vehicle models and motion states under different road conditions had important reference value for us to reduce vehicle emissions and energy consumption.

In order to effectively measure the behavioural changes on exhaust emission. Many researchers had put forward various traffic flow models theoretically, such as traffic flow mechanics model, dynamic theory model, following model and cellular automata model cet. [2,9,10,11] Cellular automata models (CA) were discrete models of time, space and state variables. Its algorithm was simple and easy to calculate synchronously, which was suitable for large-scale computer simulation and could effectively simulate the motion state of road traffic vehicles. Therefore, cellular automata model has been widely studied and applied in traffic flow.[2,6,12] The most famous cellular automata traffic flow model was the random NaSch model proposed by Nagel and Schreckenberg in 1992. [13] On the basis of NaSch model, many improved models have been proposed, such as TT, BJH, VDR and FI model [14-17] Wang Bing-Hong et al.[18] determined the velocity at the next step through the workshop distance, and successfully obtained the analytical solution of the FI model. Tian Huan-Huan et al.[19] studied the energy consumption of the NaSch model of cellular automata mixed traffic flow by considering the different influences of vehicle length, maximum velocity and delay probability. Pan et al.[20] have resulted fuel rate, dissipation component and particle emission on speed limit, injection rate and extinction rate of the lane of Traffic emitted particle modeling based on NaSch Model of a single lane. Shi Wei et al. [21] analyzed the relationship between the stability and additional energy consumption of several typical galloping models. Wang Tao[22] and Toledo et al.[23] studied the energy consumption of galloping models and urban traffic models respectively. Nan Cui [24] studied the effects of different route guidance strategies on traffic emissions in intelligent transportation systems. Results showed that mean velocity route guidance strategy and congestion coefficient route guidance strategy were the best ones. They can effectively reduce the traffic emissions and improve the efficiency of traffic network. Wang Xue et al. [25] simulated the emissions of mixed traffic by the NaSch model of cellular automata, and found that the vehicles for a pattern of stop-go movement emitted a large number of particulate emission. Julio César Pérez-Sansalvador [26] used a cellular automata and an instantaneous traffic emissions model to capture the effect of speed humps on traffic flow and on the generation of $CO_2$, $NO_x$, VOC and PM pollutants. They found that speed humps significantly promote the generation of pollutants when the number of vehicles on a lane is low. N. Lakouari [27] proposed that traffic control at intersections in urban conditions directly influences the $CO_2$ emissions in their proximity. Improving traffic control at an intersection is another way to reduce vehicular emissions and creating a positive impact on the environment.

---
[*] Corresponding author: Dong Chen (chendong1418@163.com)

This paper studies the energy consumption and PM emission of cellular automata FI model of mixed traffic flow. Study the influence of maximum velocity, mixing ratio and random deceleration probability on traffic energy consumption and PM emission, explore the characteristics energy consumption and emission in FI model, and look for the traffic model with low energy consumption and emission.

## 2. Methodology
2.1 Mixed Fukui-Ishibashi (FI) model

Cellular automata models (CA) were discrete models of time, space and state variables. Its algorithm was simple and easy to calculate synchronously, which was suitable for large-scale computer simulation and could effectively simulate the motion state of road traffic vehicles. Therefore, cellular automata model has been widely used in the study of traffic flow. The most famous cellular automata traffic flow model was the random NaSch model proposed by Nagel and Schreckenberg in 1992. Later, many scholars proposed many improved models on the basis of NaSch model, such as TT BJH VDR and FI model. Vehicles will generate a large amount of exhaust gas and energy consumption in the stop-and-go state, when the vehicle is at the intersection, the phenomenon of stop-and-go is especially obvious. FI model takes crossroads as the research object. Therefore, this paper chooses FI model as traffic model.

In the mixed FI model, the road is assumed to be a one-dimensional discrete closed grid with a growth degree of L in mixed FI model composed of vehicles with different lengths and every moment each cell may be occupied or without vehicles. At each evolutionary time step $t \rightarrow t+1$, all vehicle states are updated in parallel according to the evolution rules of FI model. The update rules of each vehicle state evolution in mixed traffic flow are described as follow.

Step 1. Velocity update

If $(v_{max} \leq gap(i,t))$, then

$$v(i,t+1) = \begin{cases} v_{max} & 1-p \\ v_{max}-1 & p \end{cases} \quad (1)$$

Else

$$v(i,t+1) = gap(i,t) \quad (2)$$

Step 2. Traffic motion

$$x(i,t+1) \rightarrow x(i,t) + v(i,t+1) \quad (3)$$

Where $v(i,t) \in [0, v_{max}] (i \in [1, N])$ is the velocity of the $i$-th vehicle at time $t$. $v_{max}$ is the maximum velocity, and $N$ is total number of vehicles. $x(i,t)$ is the position of the $i$-th vehicle at time $t$. The gap between $i$-th vehicle and $(i+1)$-th vehicle denotes $gap(i,t) = x(i+1,t) - x(i,t) - Length(i)$. $Length(i)$ is the length of the $i$-th vehicle. When the probability $p$ is 0 or 1, we consider the FI model is deterministic.

### 2.2 Emission model

Panis et al. used emission functions for instantaneous velocity and acceleration as each vehicle parameters using non-linear multiple regression techniques to establish an emission model. $E(i,t)$ is pollutant emission for the i-th vehicle in unit time.

$$E(i,t) = \max[E_0, f_1 + f_2 v(i,t) + f_3 v(i,t)^2 + f_4 a(i,t) + f_5 a(i,t)^2 + f_6 v(i,t)a(i,t)] \quad (4)$$

where $E_0$ is a lower limit of emission (g/s) specified for each vehicle and pollutant type, and $f_1$ to $f_6$ are emission parameter specific for each vehicle and pollutant type determined by the regression analysis. All of specified parameters value is given in Table 1.

In order to determine overall vehicle emissions, the average vehicle emission based on equation (4) is defined as follows.

$$\bar{E} = \frac{1}{T}\frac{1}{N} \sum_{t=t_0+1}^{t_0+T} \sum_{i=1}^{N} E(i,t) \quad (5)$$

where $E(i,t)$ represents the pollutant emission of the $i$-th vehicle at time $t$. The average vehicle emission reflects pollutant emissions of traffic flow on the whole. $N$ is the total number of vehicles on the road, and $t_0$ is the relaxation time. We set $E_d(i,t)$ is deterministic emission, which is caused by the determination of deceleration between vehicles. $E_r(i,t)$ represents random emission which is caused by random deceleration. So, the total average emission $E(i,t) = E_d(i,t) + E_r(i,t)$.

**Table 1** Parameters of emission model

| Pollutant | Vehicle type | $E_0$ | $f_1$ | $f_2$ | $f_3$ | $f_4$ | $f_5$ | $f_6$ |
|---|---|---|---|---|---|---|---|---|
| PM | Diesel car | 0 | 0.00e+00 | 3.13e-04 | -1.84e-05 | 0.0e+00 | 7.50e-04 | 3.78e-04 |
|  | Bus | 0 | 2.23e-04 | 3.47e-04 | -2.38e-05 | 2.08e-03 | 1.76e-03 | 2.23e-04 |

**2.3 Energy dissipation model**

The kinetic energy of the vehicle is defined as $1/2 mv^2$ with the velocity $v(i,t)$. Where $m = \text{lenght}(i) * m_0$ is the mass of the vehicle and $m_0$ is the mass when the length is one cell. When the vehicle decelerates, the kinetic energy of the vehicle decreases. $D(i,t)$ represents the average energy consumption of each vehicle per unit time and defines the energy consumption of the $i$-th vehicle from time $t$ to $t+1$ as:

$$D(i,t) = \begin{cases} \frac{m}{2}[v^2(i,t) - v^2(i,t+1)], & v(i,t+1) < v(i,t) \\ 0 & v(i,t+1) \geq v(i,t) \end{cases} \quad (6)$$

In order to determine overall vehicle energy dissipations, the average vehicle energy dissipation based on equation (4) is defined as follows.

$$\bar{D} = \frac{1}{T}\frac{1}{N} \sum_{t=t_0+1}^{t_0+T} \sum_{i=1}^{N} D(i,t) \quad (7)$$

Where determined energy consumption $D_d(i,t)$ represents determined energy consumption caused by deceleration and random energy consumption $D_r(i,t)$ represents energy consumption caused by random deceleration then the total average energy consumption $D(i,t) = D_d(i,t) + D_r(i,t)$.

## 3. Simulations and discussions

Periodic boundary condition is used to the simulation, selecting the road length are $L = 10^4$ cells. The average vehicle emission, the average vehicle energy consumption and the averages velocity are obtained in the process of simulations by averaging over 50 independent initial realizations up to $3 \times 10^5$ iteration steps for each run and by discarding the first $2 \times 10^5$ iteration steps as transient time, the global density is $\rho = N/L$. Therefore, the occupancy of the lane is expressed as $C = N \times \text{Length}/L$, it is the ratio of the number of cells occupied by the vehicle to the length of the lane. The number of long vehicles is denoted as NL and correspondingly the number of short one is expressed as NS. Long vehicle occupies LL cells and short one occupies a unit cell. The occupancy of long vehicles is the ratio of the number of long vehicle times its length to the length of lane ($C^l = N^l \times L^l / L$) and accordingly the occupancy of short one equals to $C^s = N^s \times L^s / L$. The mixing ratio $C_{\text{mix}}$ is defined as $C_{\text{mix}} = C^l / C$, which is to measure the proportion of long vehicles in mixed traffic flow.18 The mass of the single vehicle is denoted as $m=1$.

The average velocity $\bar{v}$ of traffic flow is defined as follows.

$$\bar{v} = \frac{1}{T} \frac{1}{N} \sum_{t=t_0+1}^{t_0+T} \sum_{i=1}^{N} v(i,t) \tag{8}$$

### 3.1 The impact of fuel consumption and PM emission for different motion state

In order to investigate the impact of fuel consumption and PM emission, we firstly analyze the fundamental diagram of mixed traffic flow with different mixing ratios (flow and occupancy rate). In Figure 1, we set the short vehicle is 1 cell and the long vehicle 10 cell. The maximum speed is set to $v_{\text{max}} = 5$ cell/s. The fundamental diagram of different mixed ratios ($C_{\text{mix}}$) is obtained. In Figure 1, we could find when $C_{\text{mix}}=0$ (all vehicles are short), the flow rises rapidly to the peak and the peak of global flow is the largest. When $C_{\text{mix}}=1$ (all vehicles are long), the flow slowly rises to the peak and the peak of global flow is the smallest. With the increase of vehicle mixed ratios, the peak of global flow gradually moves to the right and the peak decreases.

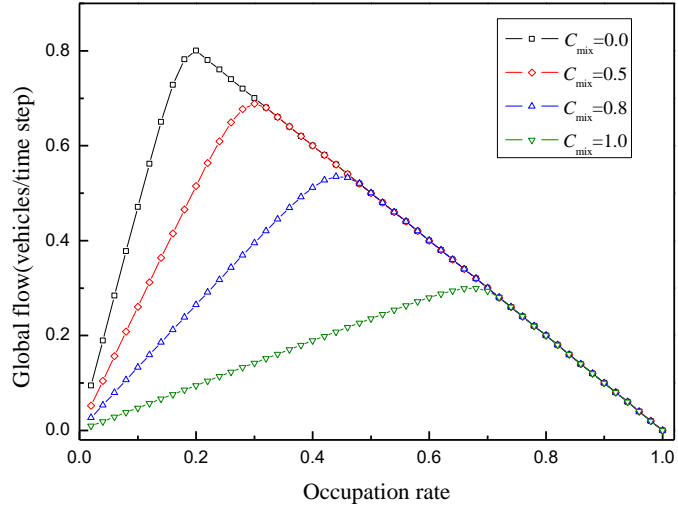

Figure 1 The fundamental diagram of different mixed ratios ($p=0.25$)

When the random deceleration probability $0<p<1$, FI model is non-deterministic model. Thus, the average PM emission is $E(i,t) = E_d(i,t) + E_r(i,t)$ and the average energy consumption is $D(i,t) = D_d(i,t) + D_r(i,t)$. Figure 2 shows the relationship between global flow and occupancy rate when the random slowing probability $p=0.25$ and $v_{\text{max}} = 2,5,8,10$. In Figure 2, the maximum peak of global flow is about 0.52 when the occupation rate $O \approx 0.41$ for $v_{\text{max}} = 2$. The maximum peak of global flow is about 0.8 when the occupation rate $O \approx 0.21$ for $v_{\text{max}} = 5$. The maximum peak of global flow is about 0.86 when the occupation rate $O \approx 0.13$ for $v_{\text{max}} = 8$. The maximum peak of global flow is about 0.9 when the occupation rate $O \approx 0.1$ for $v_{\text{max}} = 10$. The simulation experiment analysis indicates that the higher the maximum speed, the greater the peak of global flow. Meanwhile, the peak of global flow moves to the left continuously and rises rapidly, as the maximum speed increases.

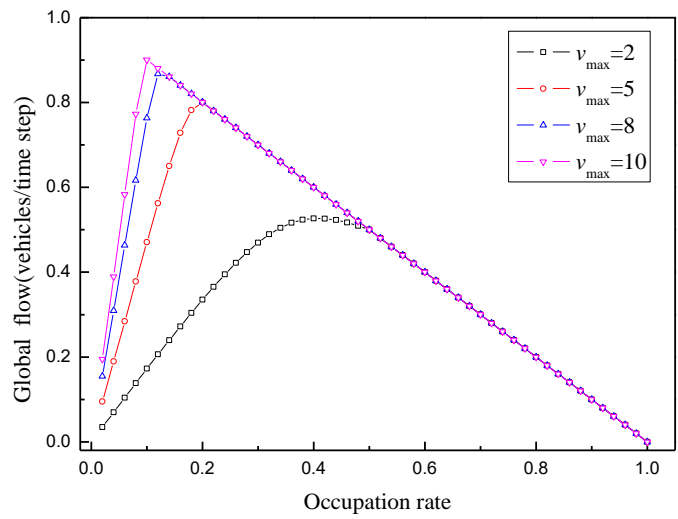

Figure 2 The fundamental diagram of different max velocity (*p*=0.25)

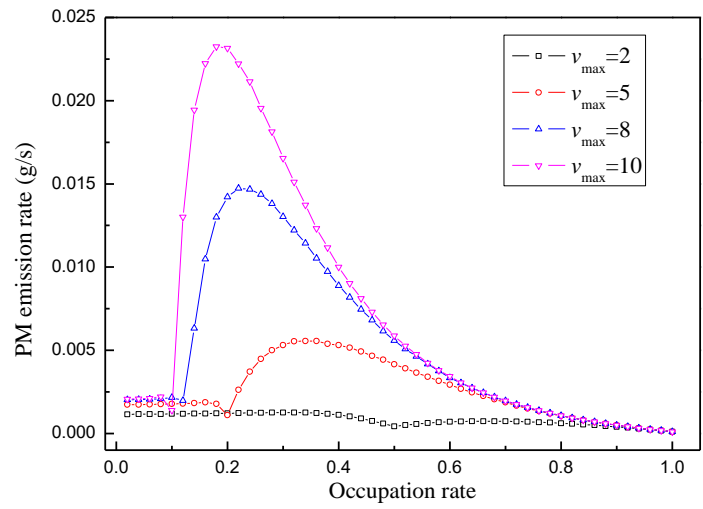

(a)

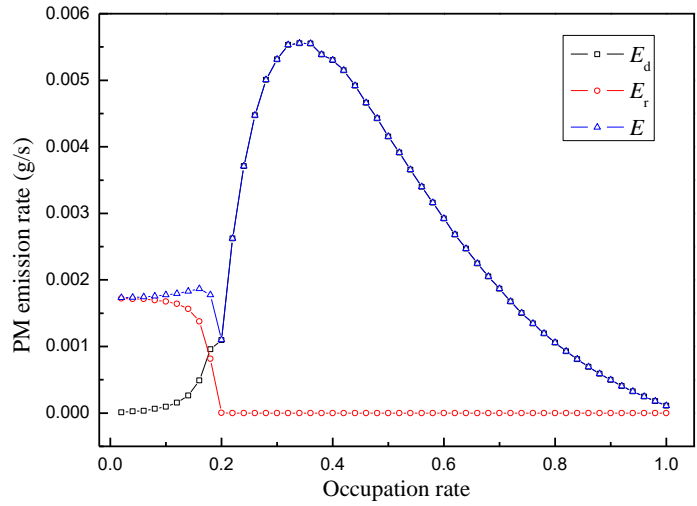

(b)

Figure 3 (a) The relationship between PM emission rate and occupation rate for different maximum speeds and $C_{mix}$=0.2; (b) The relationship between PM emission rate and occupation rate for $v_{max}$=5 and $C_{mix}$=0.2

Figure 3 (a) shows the particulate matter emission rate for different maximum speeds $v_{max} = 2, 5, 8, 10$. When $v_{max} = 2$, the PM emission is very little and is no big fluctuation, which indicates that the smaller the maximum speed, the lower the PM emission. Meanwhile, the variation trend of PM emission in Figure 3 (a) corresponds to the variation trend of global flow in Figure 2. In Figure 3 (b), we discuss the relationship between PM emission rate and occupation rate for $v_{max}$=5 and $C_{mix}$=0.2. When global traffic is at its peak, the probability of random deceleration is 0. Thus, PM emissions are generated only by the vehicles with deterministic deceleration. When the occupation rate is $O$≈0.38, the PM emission reaches the zenith.

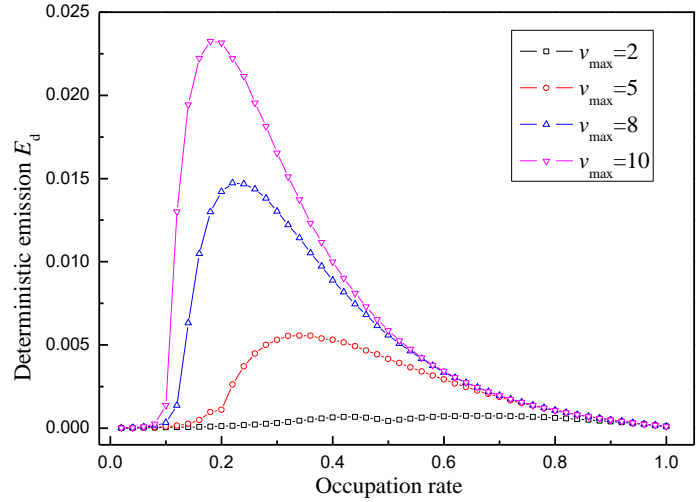

(a)

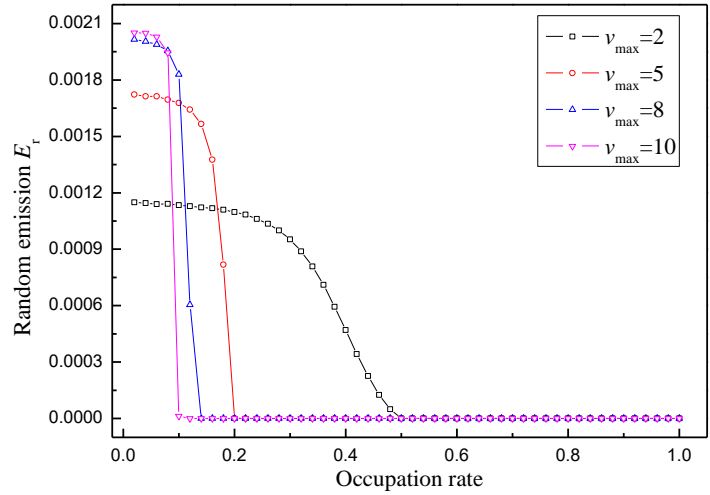

(b)

Figure 4 (a) the relationship between deterministic PM emission rate and occupancy rate for different maximum speeds; (b) the relationship between random PM emission rate and occupancy rate for different maximum speeds

The deterministic and stochastic PM emissions for different maximum speeds are shown respectively in Figure 4 (a) and (b). In Figure 4 (a), when the max speed is $v_{max}=2$, the deterministic PM emission reached the peak at occupancy $O\approx 0.42$. When the max speed is $v_{max}=5$, the deterministic PM emission reached the peak at occupancy $O\approx 0.31$. For $v_{max}=8$, the peak is at occupation rate $O\approx 0.21$. For $v_{max}=10$, the peak is at occupation rate $O\approx 0.19$. The results show that the peak of deterministic PM emission increases with the increase of max speed and moves to the left gradually. Moreover, the deterministic PM emission starts at 0 and rapidly increases to the peak and then slowly decreases. In Figure 4 (b), the random PM emission $E_r$ drops slowly before locating the maximum and then suddenly drops to zero. Meanwhile, we observe that the higher the maximum speed, the faster the descent.

In Figure 5 (a), the relationship between energy consumption and occupancy rate for different maximum speeds is shown. In the free flow region corresponding to the fundamental diagram (Figure 2), the energy consumption is very low, but increases slowly with the increase of occupancy rate. When the total flow reaches the maximum in Figure 2, the energy consumption suddenly drops to the minimum, then rises dramatically to the maximum, and then drops gently. With the increase of the maximum speeds, the maximum energy consumption increases and the peak of energy consumption moves to the left. To further explain this phenomenon, we take $v_{max}=5$ as an example for analysis in Figure 5 (b). When occupancy rate is $O\approx 0.2$, traffic energy consumption drops sharply at the maximum flow. The reason is that after this point, the space between vehicles is less than the maximum speed of vehicles. So, the random energy consumption is reduced to 0, which is provided by deterministic vehicles.

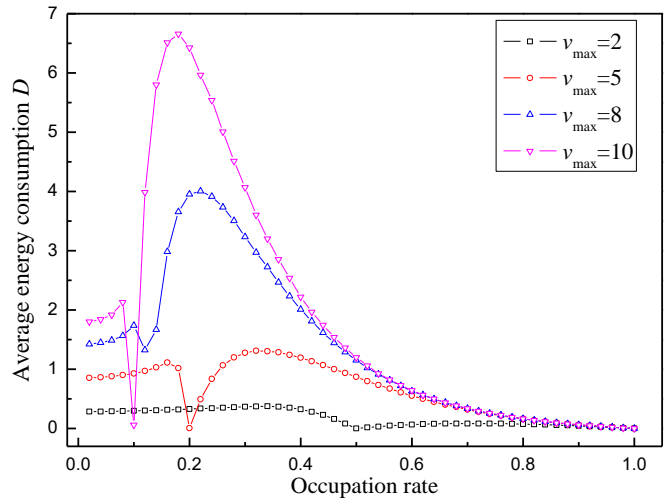

(a)

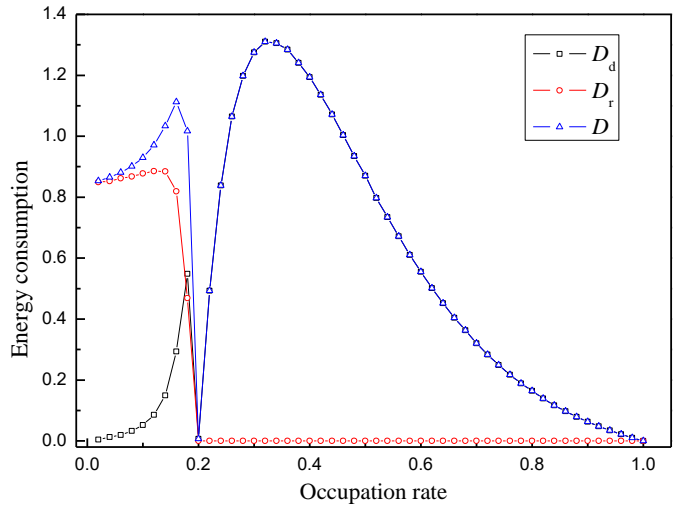

(b)

Figure 5 (a) the relationship between energy consumption and occupancy rate for different maximum speeds; (b) The relationship between energy consumption and occupation rate for $v_{max}=5$ and $C_{mix}=0.2$

In Figure 6 (a) and (b), the energy consumption of deterministic energy consumption $D_d$ and random energy consumption $D_r$ are also given for different maximum speeds. In Figure 6 (a), the energy consumption corresponding to different maximum speeds increases quickly, then decay slowly. In Figure 6 (b) is the random energy consumption corresponding to different maximum speeds. It increases from $O=0.0$ and has the maximum energy consumption at the maximum flow rate, and then suddenly drops to zero.

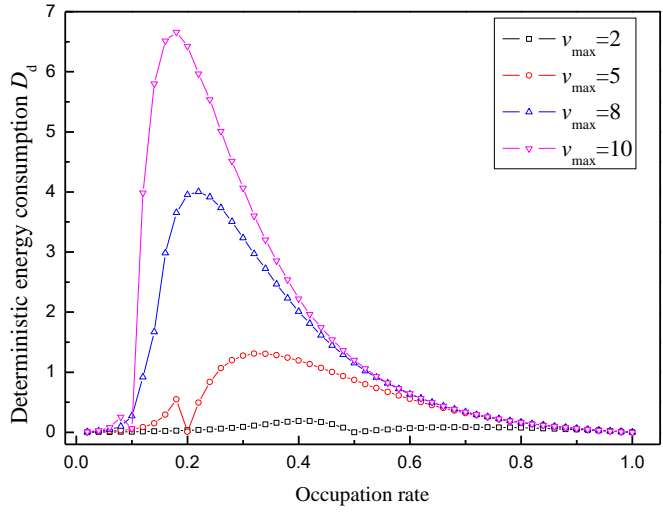

(a)

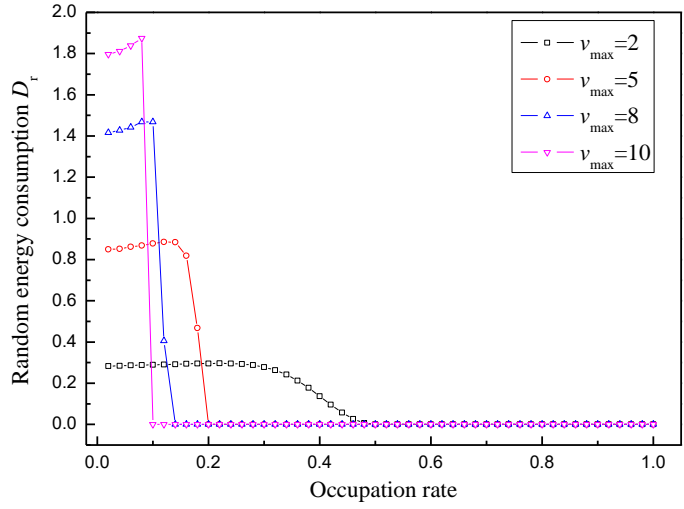

(b)

Figure 6 (a) the relationship between deterministic energy consumption and occupancy rate for different maximum speeds; (b) the relationship between random energy consumption and occupancy for different maximum speeds

In order to further study the characteristics of emission and energy consumption of FI model, the fraction of random deceleration and deterministic deceleration is introduced. We define it as

$$f_{r(d)} = \frac{1}{T}\frac{1}{N}\sum_{t=t_0+1}^{t_0+T} n_{r(d)}(t) \qquad (9)$$

Where $n_{r(d)}(t)$ is the total number of vehicles with random (deterministic) deceleration in time $T$. The fraction of vehicles with random (deterministic) deceleration can be obtained by numerical simulation in Figure 7. As can be seen from the figure 7, at the occupancy rate of $O=0.2$, there is no random deceleration of vehicles, and the corresponding vehicle does not deterministic deceleration, so that the PM emission and the energy consumption of traffic flow is minimal. After $O>0.2$, only the decelerated vehicle has an effect and starts to increase and then decrease, making the change of PM emission and energy consumption corresponding to Figure 4(a) and Figure 6(a).

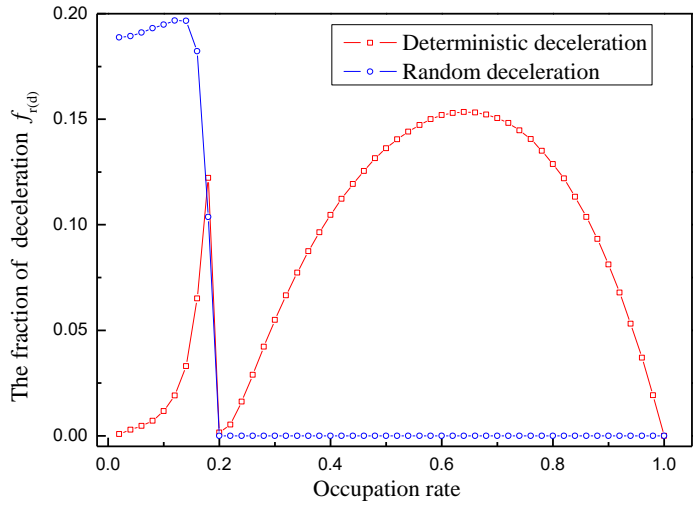

Figure 7 The fraction of random and deterministic deceleration vehicles for $v_{max}=5$

## 3.2 The impact of mixed ratio on PM emission and energy consumption

Figure 8 shows the relationship between occupation rate and global flow for different mixing ratios. It can be seen from Figure 8 that the global flow decreases as the mixing ratio increases, and the maximum flow value moves to the right with the increase of mixing ratio. The flow increases continuously with the increase of the share and reaches the maximum value and then rapidly decreases to 0.

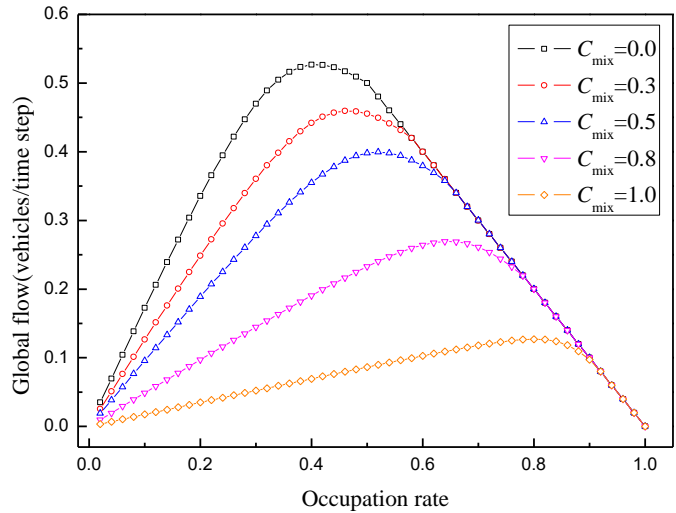

Figure 8 The relationship between occupation rate and global flow for different mixing ratios when $p=0.25$ $v_{max}=2$

In order to study the influence of mixing ratio on PM emission and energy consumption of mixed traffic flow composed of vehicles of different lengths, this part is the mixed traffic flow composed of vehicles with short vehicle Length_I=1, long vehicle Length_II=10, maximum velocity of 2 and random deceleration probability $p=0.25$. Figure 9 (a) shows the relationship between total average PM emission and occupancy at different mixing ratios $C_{mix}$. With the increase of occupancy, the average total emission first slowly increases to the maximum value, then decreases rapidly, and finally decreases to a value close to 0. The higher the mixing ratio, the higher the average emission, and the maximum emission keeps moving to the right. Figure 9 (b) shows the situation of determined emission, random emission and average total emission when the mixing ratio is 0.5. The random emission decreases with the increase of the occupation rate, and finally decreases to 0. With the increase of the occupancy rate, the determined emission first increases continuously. When the emission is determined to reach a maximum, it suddenly decreases, then continues to rise to a larger value and finally decreases to a certain value.

Figure 10 (a) and (b) show the determined and random PM emissions under different mixing ratios respectively. Figure 10 (a) shows that the larger the mixing ratio is, the larger the maximum determined emission will be and gradually shift to the right. Figure 10 (b) shows that with the increase of the mixing ratio, the PM emission keeps increasing and the maximum random emission keeps moving to the right. As the occupancy rate increases, random emissions keep unchanged at first and then decrease to 0.

Figure 11 (a) is the change diagram of total average energy consumption $D$ and occupancy rate at different mixing ratios $C_{mix}$. The total average energy consumption slowly increases to the maximum value with the increase of occupancy rate. After entering the congestion area, the total average energy consumption suddenly decreases to 0, then gradually increases to a maximum value with the increase of occupancy rate $O$, and finally decreases to 0. The energy consumption generated by random motion in this stage is 0, and the average total energy consumption is equal to the energy consumption generated by the decelerated motion. With the increase of mixing ratio $C_{mix}$, the occupancy rate moves to the right when the total average energy consumption reaches the maximum value. Figure 11 (b) shows the changing relationship between mixing ratio $C_{mix}$ =0.5 and energy consumption $D$, $D_d$, $D_r$ and occupancy rate.

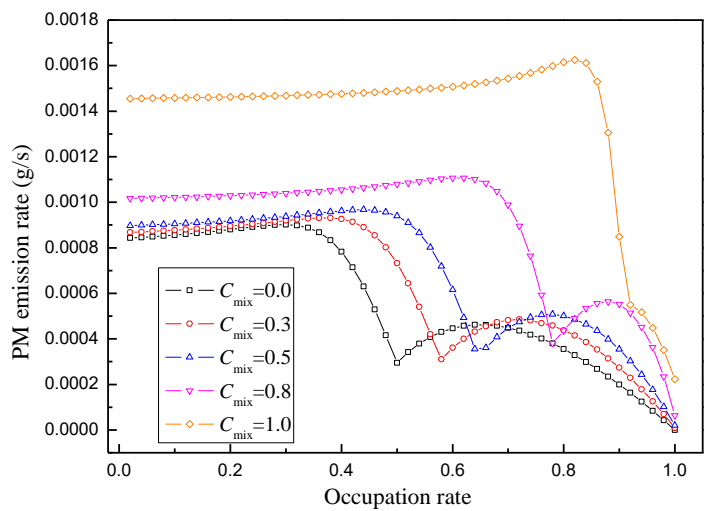

(a)

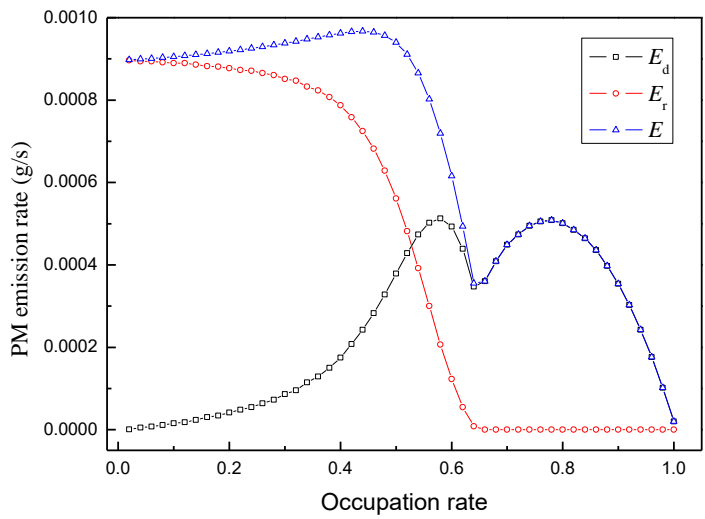

(b)

Figure 9 (a) The relationship between particulate emissions and occupation rate for different mixing ratios $C_{mix}$; (b) The relationship between particulate emission and occupation rate for $C_{mix}=0.5$

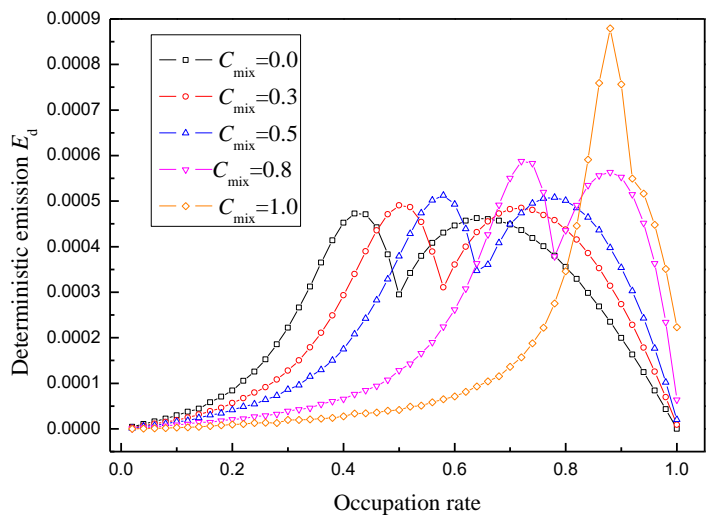

(a)

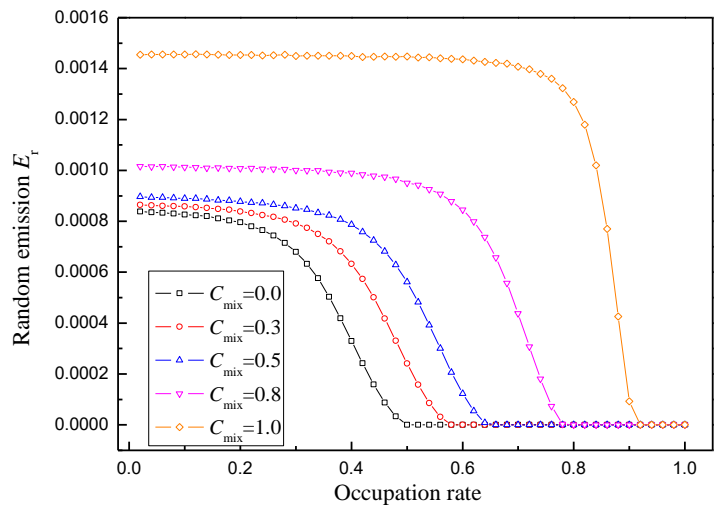

(b)

Figure 10 The relationship between particulate emissions and occupation rate for different mixing ratios $C_{mix}$

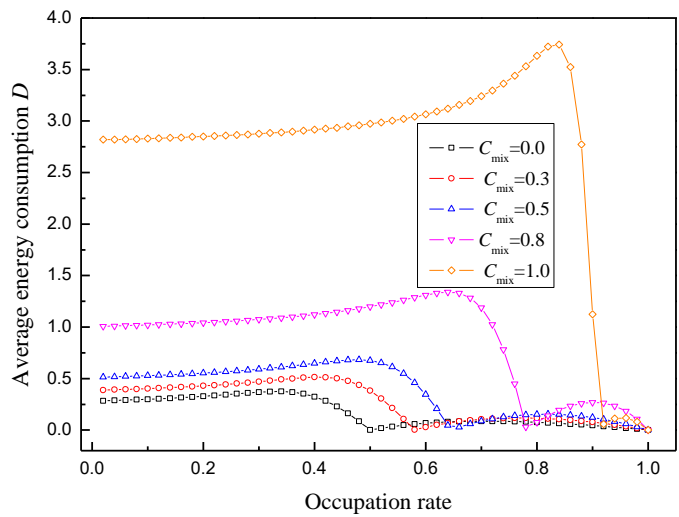

(a)

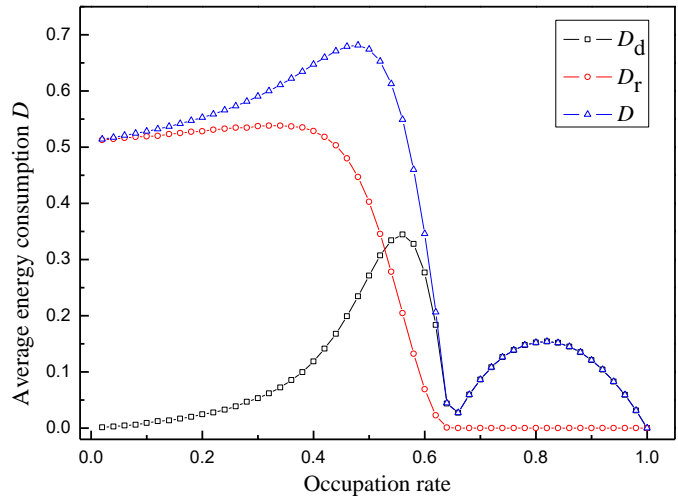

(b)

Figure 11 (a) The relationship between average energy consumption and occupation rate for different mixing ratios $C_{mix}$; (b) The relationship between average energy consumption and occupatioon rate for $C_{mix}=5$

Figure 12 shows the relationship between the deterministic energy consumption $D_d$, $D_r$ and occupation rate with different mixing ratios. Figure 12 (a) shows that the energy consumption $D_d$ increases to the maximum then rapidly decreases to a value approaching 0, gradually increases to a maximum point, and finally decreases to 0. With the increase of mixing ratio $C_{mix}$, the occupancy rate shifted to the right when the energy consumption $D_d$ reached the maximum value. Figure 12 (b) shows that $D_r$ of random energy consumption decreases slowly to 0 with the increase of occupation rate then remains unchanged. With the increase of mixture ratios, $D_r$ of energy consumption decreases to 0.

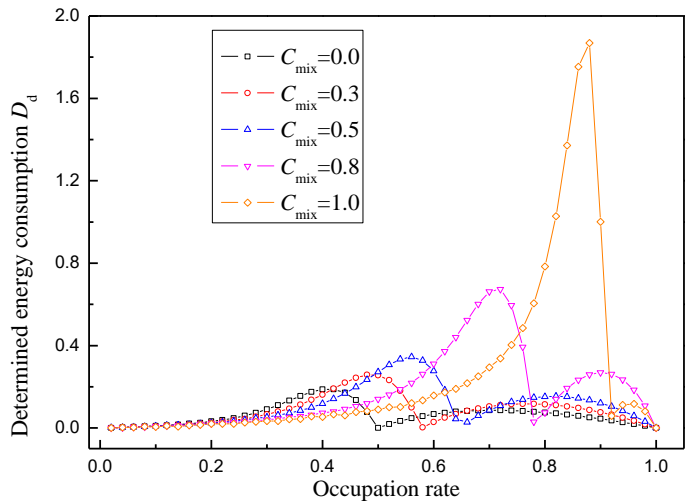

(a)

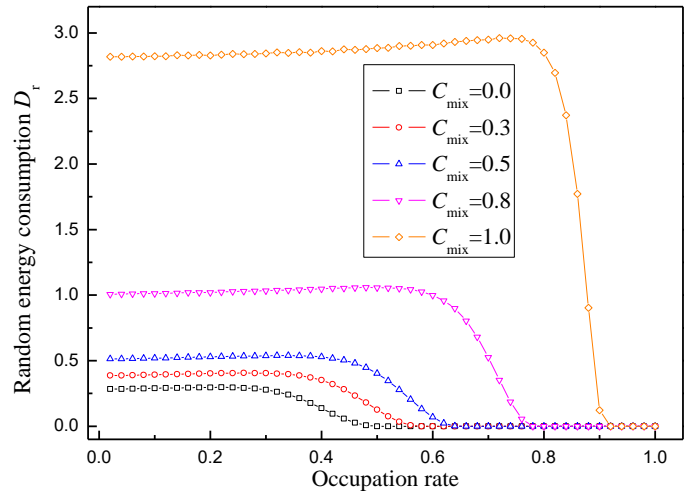

(b)

Figure 12 (a) the relationship between deterministic energy consumption and occupancy rate different mixing ratios $C_{mix}$; (b) the relationship between random energy consumption and occupancy different mixing ratios $C_{mix}$

## 4. Conclusions

In this paper, the influences of maximum velocity, mixing ratio and random deceleration probability on energy consumption and PM emission of FI traffic flow model under periodic boundary conditions are studied. The numerical simulation results show that, when the vehicle length is constant, the energy consumption and PM emission of the FI traffic flow model increase significantly with the increase of the maximum velocity of the vehicle. In the mixed traffic flow with the same maximum velocity and different vehicle lengths, the energy consumption and PM emission of the FI traffic flow model increase with the increase of the mixing ratio. For the FI traffic flow model, the sudden drop tends to 0 in traffic energy consumption at the maximum global flow, and the PM emission also drops suddenly at the maximum flow rate. There are maxima of energy consumption and emission around the maximum global flow. The maximum velocity and length of the vehicle are fixed, and the random deceleration probability p is different. When the energy consumption and emission of FI are larger than the critical occupancy, the emission and energy consumption curves overlap respectively.

## Acknowledgments

This work was supported by the Doctoral Research Project of Guangxi University of Finance and Economics (Grant No. BS2019027), Project of Improving the Basic Scientific Research Ability of Young and Middle-Aged Teachers in Guangxi Universities (Grant No. 2020KY16023). In addition, this research was funded by the management science and engineering discipline construction fund of Guangxi University of Finance and Economics.